\documentclass[fleqn,10pt]{wlscirep}
\usepackage[utf8]{inputenc}
\usepackage[T1]{fontenc}
\usepackage{bm}
\usepackage{graphicx}
\usepackage{multirow}
\usepackage{amsmath,amssymb,amsfonts}
\usepackage{amsthm}
\usepackage{mathrsfs}
\usepackage{xcolor}
\usepackage{textcomp}
\usepackage{manyfoot}
\usepackage{booktabs}
\usepackage{algorithm}
\usepackage{algorithmicx}
\usepackage{algpseudocode}
\usepackage{listings}

\title{From Pixels to Camera: Scaling Superconducting Nanowire Single-Photon Detectors for Imaging at the Quantum-Limit}

\author[1]{Jun Gao}
\author[2]{Jin Chang}
\author[3]{Bruno Lopez Rodriguez} 
\author[3]{Iman Esmaeil Zadeh} 
\author[1]{Val Zwiller}
\author[1,*]{Ali W. Elshaari}

\affil[1]{Department of Applied Physics, KTH Royal Institute of Technology, Albanova University Centre, Stockholm, SE-106 91, Sweden}
\affil[2]{\protect\raggedright Kavli Institute of Nanoscience, Department of Quantum Nanoscience, Delft University of Technology, 2628CJ Delft, the Netherlands.}

\affil[3]{\protect\raggedright 
Department of Imaging Physics (ImPhys), Faculty of Applied Sciences, Delft University of Technology, Delft 2628 CJ, The Netherlands}

\affil[*]{e-mail: elshaari@kth.se}

\begin{abstract}
Superconducting nanowire single-photon detectors (SNSPDs) have emerged as essential devices that push the boundaries of photon detection with unprecedented sensitivity, ultrahigh timing precision, and broad spectral response. Recent advancements in materials engineering, superconducting electronics integration, and cryogenic system design are enabling the evolution of SNSPDs from single-pixel detectors toward scalable arrays and large-format single-photon time tagging cameras. This perspective article surveys the rapidly evolving technological landscape underpinning this transition, focusing on innovative superconducting materials, advanced multiplexed read-out schemes, and emerging cryo-compatible electronics. We highlight how these developments are set to profoundly impact diverse applications, including quantum communication networks, deep-tissue biomedical imaging, single-molecule spectroscopy, remote sensing with unprecedented resolution, and the detection of elusive dark matter signals. By critically discussing both current challenges and promising solutions, we aim to articulate a clear, coherent vision for the next generation of SNSPD-based quantum imaging systems. 
\end{abstract}

\begin{document}

\flushbottom
\maketitle

\thispagestyle{empty}

\section*{Introduction: Toward Large-Scale Quantum Imaging Systems}

Superconducting nanowire single-photon detectors (SNSPDs) have emerged over the past two decades as revolutionary photon-counting devices, offering performance capabilities unmatched by those of traditional semiconductor-based photodetectors. Their exceptional detection efficiency, ultrahigh timing resolution in the picosecond regime, negligible dark counts, and broad spectral response ranging from X-ray to far-infrared wavelengths \cite{guo2024high,taylor2023low} have positioned them as critical enabling components in diverse scientific and technological domains \cite{esmaeil2021superconducting,chang2023nanowire}. Originally designed as single-pixel detectors for quantum optics and quantum information experiments\cite{you2020superconducting}, recent progress in material engineering, nanofabrication methods, and cryogenic electronics has driven the development of SNSPDs toward multi-pixel arrays\cite{steinhauer2021progress}, enabling entirely new classes of quantum-limited imaging and sensing applications\cite{defienne2024advances}. The scaling of SNSPD arrays into large-format cameras promises transformative impact across multiple fields, from quantum communication networks requiring secure high-speed single-photon detection, to deep-space optical sensing\cite{hao2024compact} and biomedical imaging techniques capable of visualizing molecular-level biological processes at extremely low photon fluxes\cite{tamimi2024deep,wang2021non} and with unprecedented time resolution on every photon detection event. Quantum communication, for example, critically depends on the reliable detection of individual photons over long distances with minimal timing uncertainty-demands that are ideally suited to SNSPD technology. While standard QKD protocols typically use single-pixel detectors, emerging quantum applications, such as high-dimensional encoding, may benefit from SNSPD arrays or camera technologies where different modes and photon numbers can be resolved with multipixel detectors. In addition, long distance free-space communication (e.g. ground station satellite communication), to cope with the large and distorted beams (due to air turbulence), requires large area detectors which still have high performance metrics in terms of efficiency and speed, currently attainable only in array detectors. Similarly, applications such as remote sensing and astronomical imaging require detectors with superior sensitivity and low noise, both of which are naturally provided by superconducting nanowire detectors across different wavelengths~\cite{chang2021detecting,chang2022efficient}. Despite their significant promise, scaling SNSPD arrays from individual pixels to integrated large arrays imaging systems introduces substantial scientific and engineering challenges\cite{steinhauer2021progress,mccaughan2018readout}. Some of these are stringent cryogenic operation requirements, complex thermal management issues, and the increasing complexity of read-out electronics, along with the handling of large data of detection events. Conventional SNSPDs operate below 4 K, and in the case of common amorphous superconductors (e.g. a-WSi and a-MoSi) in the sub-Kelvin regimes. At these low temperatures, the thermal cooling budget is limited, and even minimal heat dissipation through wiring, amplification, or biasing circuitry can adversely affect the performance of the entire detector array. This constraint severely limits traditional approaches to directly connect each pixel with individual bias and read-out coaxial RF cables, as the thermal load and mechanical complexity rapidly become prohibitive when scaling beyond small number of pixels. To address these limitations, researchers have developed several advanced multiplexing and read-out strategies, including row-column multiplexing, inductor-based pulse shaping, frequency division multiplexing, superconducting single-flux quantum (SFQ) digital electronics integration, thermally-coupled pixel arrays, and compressive sensing-based single-channel read-out architectures \cite{wollman2019kilopixel,mccaughan2022thermally,allmaras2020demonstration,zhang202332,miyajima2019single,miyajima2018high,yabuno2020scalable,luskin2023large,guan2022snspd,oripov2023superconducting,doerner2017frequency,ortlepp2011demonstration}. Each of these strategies attempts to strike an optimal balance between maintaining high timing accuracy, minimal heat load, and scalable system complexity. For instance, row-column multiplexing significantly reduces wiring complexity by encoding spatial information through intersecting rows and columns, while SFQ electronics digitize photon detection events at cryogenic temperatures to reduce the number of interconnections and thermal dissipation to the superconducting pixels. Likewise, recent demonstrations of compressive sensing have shown how computationally intensive reconstruction algorithms can effectively decode multi-pixel photon detections from highly sparse measurement signals, offering pathways toward scalable, large-area detector arrays\cite{guan2022snspd}.

This perspective critically reviews recent advances and remaining challenges in scaling SNSPD arrays into fully integrated imaging systems. We first discuss state-of-the-art developments in superconducting materials, detector architectures, and innovative multiplexing strategies and then examine the key physical and engineering bottlenecks, including cryogenic thermal management, wiring complexity, and readout fidelity, that currently hinder large-scale array implementation. By outlining promising approaches and emerging solutions, we aim to stimulate further scientific dialogue and interdisciplinary collaboration among materials scientists, cryogenic engineers, quantum physicists, and computational imaging specialists. Our goal is not only to highlight recent progress but also to articulate clear pathways toward achieving scalable, robust SNSPD camera systems capable of transforming photon-starved imaging applications in quantum technologies, biomedical diagnostics, remote sensing, and fundamental scientific exploration.

\section*{Emerging Materials and Advanced Detector Architectures for SNSPD Arrays}

The rapid advancement of superconducting nanowire single-photon detectors (SNSPDs) has been driven largely by innovative developments in both superconducting material systems and detector architectures. Several material platforms have emerged as frontrunners, each with unique properties tailored for different operational requirements, spectral sensitivities, and fabrication processes, significantly enhancing detector performance and scalability. Amorphous Tungsten Silicide (a-WSi) is renowned for its high yield and yield sensitiviry to low energy particles. a-WSi detectors have also attracted attention for cutting-edge applications such as dark matter searches due to their ultralow noise characteristics~\cite{hochberg2022new}. The amorphous nature of the films facilitates fabrication uniformity, critical for achieving high yield and reproducibility in large arrays. Moreover, the tunable stoichiometry of WSi allows precise engineering of key parameters, such as critical temperature and kinetic inductance, enabling detectors tailored for diverse applications ranging from infrared astronomy and quantum optics to advanced spectroscopy~\cite{zhang2016characteristics,verma2021single,colangelo2022large,qin2024thermal,ma2025doping,luskin2023large}. Recent advancements in optical lithography enabled the fabrication of WSi microwire detectors capable of scaling to active areas reaching millimeter and centimeter dimensions, significantly broadening their applicability while maintaining excellent single-photon sensitivity and good timing precision~\cite{chiles2020superconducting,luskin2023large}. Molybdenum silicide (MoSi) represents another compelling amorphous material system, distinguished by a comparatively high critical temperature around 7 K. This characteristic considerably simplifies cryogenic requirements, potentially enabling integration with less demanding refrigeration systems while preserving outstanding detector performance, including high SDE and good time resolution~\cite{verma2015high}. Progress in photolithographic processes has demonstrated large-scale MoSi microwire devices, further underscoring its potential for wafer-scale foundry-compatible manufacturing, thus facilitating scalable integration into quantum photonic circuits and large-format detector arrays~\cite{luskin2023large}. Niobium nitride (NbN), historically among the earliest SNSPD materials, continues to serve critical roles, especially where timing precision is essential, such as quantum key distribution (QKD) systems and time-of-flight (TOF) LiDAR applications~\cite{korzh2020demonstration,verma2021single,pearlman2005gigahertz}. Concurrently, niobium titanium nitride (NbTiN) has emerged as an attractive alternative, offering higher operating temperatures and lower kinetic inductance than NbN, providing exceptional timing jitter in the low picosecond regime\cite{esmaeil2020efficient}, making it indispensable for timing-critical quantum and classical photonic technologies.  Such attributes enhance signal-to-noise performance and detector response speed. NbTiN arrays with excellent uniformity across large active areas have been demonstrated, validating its suitability for next-generation high-density imaging and quantum sensing systems~\cite{miki2009superconducting,miki201464}. Parallel to advancements in conventional superconducting compounds, exploratory research into two-dimensional (2D) superconductors, high-T\textsubscript{c} superconductors ~\cite{chang2023superconducting} and van der Waals heterostructures presents exciting new possibilities for SNSPD architectures. For instance, twisted bilayer graphene (TBG) at so-called ``magic angles'' exhibits exotic electronic properties, such as flat-band-induced superconductivity and correlated insulating phases. Intrinsic superconductivity observed in TBG at relatively high critical temperatures (approximately 1.7\,K) suggests novel, electrostatically tunable SNSPDs, capable of bridging fundamental condensed matter physics with quantum optical detection~\cite{cao2018unconventional}. Similarly, gate-induced superconductivity in atomically thin MoS\textsubscript{2} layers demonstrates controllable superconducting transitions down to monolayers. This tunability promises dynamic, pixel-level reconfigurability of detector sensitivity, spectral response, and operational characteristics, significantly enhancing the flexibility and adaptability of future SNSPD arrays~\cite{costanzo2016gate}. Moreover, Josephson junction-based photodetectors employing graphene or semiconducting weak links leverage quasiparticle dynamics to detect single near-infrared photons efficiently. Though structurally distinct from traditional SNSPDs, these devices highlight emerging opportunities in hybrid superconducting optoelectronics, particularly suited to low-power, high-speed interconnects in cryogenic computing and quantum information systems~\cite{walsh2021josephson}. Alongside material innovations, significant progress in geometric design and detector architectures has considerably expanded SNSPD performance and scalability. Transitioning from narrow nanowires (~100 nm) to wider microwire structures (1–3 $\mu$m) has enabled dramatic enhancements in active detection areas, reducing fabrication complexity while preserving acceptable timing performance~\cite{chiles2020superconducting,pena2025high,reddy2022broadband,charaev2020large,yang2021large,protte2022laser}. This approach is particularly promising for photon-starved applications which collect light with low spatial coherence such as astronomical observations (due to atmospheric turbulence), particle detection experiments, and diffused light bio-imaging. Furthermore, novel fractal-inspired geometries, including space-filling curves and nested wire patterns, have been employed to minimize polarization sensitivity and maximize photon absorption uniformly across large detection areas. Such designs offer crucial advantages for imaging scenarios where photon arrival direction and polarization states cannot be predetermined or controlled~\cite{chi2018fractal,hao2024high,zou2024speckle,meng2022fractal}.  Finally, significant advances in integrated optical engineering, including dielectric mirror back-reflectors have been leveraged to further optimize photon absorption efficiency and spectral selectivity. Employing top-down lithography and even three-dimensional printing techniques, these optical elements enhance detector quantum efficiency across broader spectral regions, significantly improving overall device performance~\cite{redaelli2017design,china2023highly,xiao2022superconducting,munzberg2018superconducting,xiao2023ultralow}. Collectively, the convergence of cutting-edge superconducting materials, innovative nanofabrication processes, and novel optical engineering and design approach~\cite{ji2023recent} has established a robust foundation for realizing scalable SNSPD arrays. These combined innovations set a clear trajectory toward high-performance single-photon detector arrays capable of supporting transformative quantum, biomedical, astronomical, and remote sensing applications.

\section*{Innovative Multiplexing Approaches and Advanced Read-out Techniques}

The transition from single-pixel superconducting nanowire single-photon detectors (SNSPDs) to large-scale arrays suitable for quantum imaging and photon-starved detection environments hinges critically on addressing the severe technical bottlenecks posed by read-out complexity and cryogenic wiring constraints. With array size growth, traditional read-out schemes become impractical, requiring novel multiplexing methods and advanced cryogenic electronics that preserve detection fidelity while significantly reducing complexity, heat load, and integration challenges. Among established multiplexing schemes, row-column (RC) multiplexing represents one of the most straightforward yet effective approaches. In RC multiplexing, detector pixels are organized in a grid where each pixel is addressed by a combination of orthogonal bias lines corresponding to rows and columns. Photon detection events are pinpointed by identifying temporal coincidences between signals detected on these shared lines, drastically reducing the wiring requirement from a quadratic scale (\(N^2\)) to a linear scale (\(2N\))~\cite{allmaras2020demonstration,zhang202332}. This approach has been successfully demonstrated at the kilopixel scale, achieving impressive spatial localization and timing resolution, confirming its viability for near-term deployments in applications such as quantum communication and large-scale imaging systems~\cite{oripov2023superconducting,wollman2019kilopixel}. 

\begin{figure}[htbp]
    \centering
    \includegraphics[width=\linewidth]{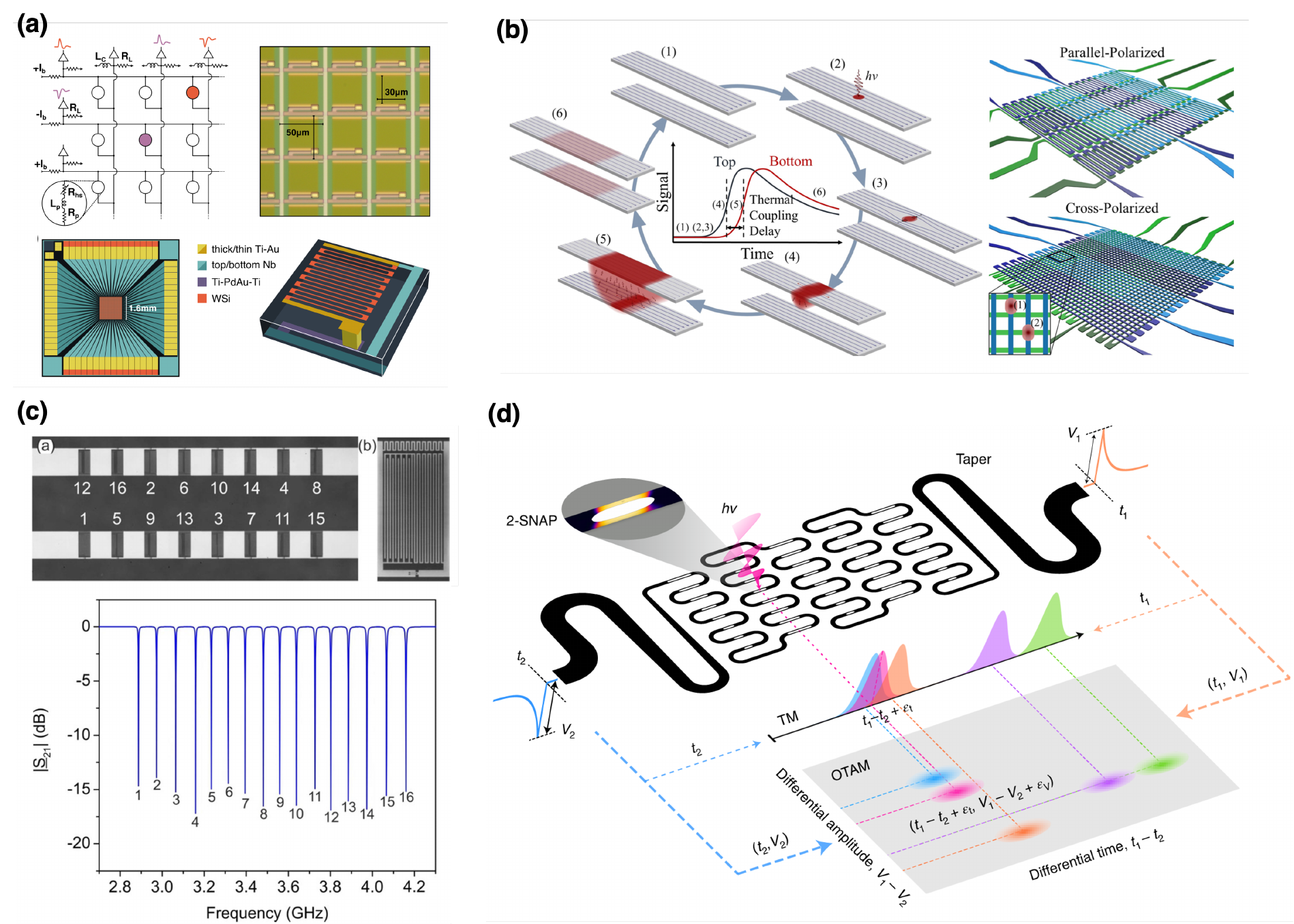}
    \captionsetup{width=\linewidth}
    \caption{Multiplexed Readout Architectures I: Selected strategies demonstrating the evolution of superconducting nanowire single-photon detector (SNSPD) array scalability.
    (a) Kilopixel SNSPD array utilizing row-column multiplexing with cryogenic readout electronics~\cite{wollman2019kilopixel}.
    (b) Thermally coupled row-column SNSPD imaging array, leveraging thermal diffusion for spatial encoding and reduced wiring overhead~\cite{allmaras2020demonstration}.
    (c) Frequency-division multiplexing where detectors are coupled to distinct superconducting resonators for shared-line readout~\cite{doerner2017frequency}.
    (d) Time–amplitude multiplexing (TAM) based on hotspot quantization, allowing pixel-specific encoding in time and amplitude domains with only two readout lines~\cite{kong2023readout}.
    }
    \label{fig:multiplexing1}
\end{figure}

Nevertheless, as the pixel count increases, RC multiplexing encounters challenges related to crosstalk, bias current redistribution, and degraded timing accuracy due to the inherent impedance and parasitic capacitance associated with longer wiring runs. Superconducting single-flux quantum (SFQ) digital processing techniques have emerged as a possible alternative. Integrating SFQ circuits directly within the cryogenic environment enables immediate digitization of photon detection signals into ultrafast, quantized voltage pulses. SFQ logic thus provides significant advantages, including dramatically reduced thermal loading, suppression of electromagnetic interference, and sub-80 picosecond timing resolution, all essential attributes for quantum information processing, high-speed photon counting, and precision metrology~\cite{yabuno2020scalable,miyajima2023single}.Building upon SFQ technology, hybrid interfaces that combine adiabatic quantum-flux-parametron (AQFP) circuits with rapid single-flux-quantum (RSFQ) logic have also been developed. These dual-mode AQFP/RSFQ architectures efficiently encode spatial and temporal photon event information using only two cryogenic transmission lines. The AQFP logic provides ultra-low-power spatial encoding through adiabatic fluxon manipulation, while RSFQ circuits maintain exceptionally high temporal precision. Experimental demonstrations with small-scale arrays have confirmed the efficacy of this hybrid approach, validating it as a promising strategy for scaling to large-format arrays with minimal cryogenic overhead~\cite{takeuchi2020scalable}. Moreover, orthogonal time-amplitude-multiplexing (TAM) method was invistigated, which represents a substantial departure from conventional electronics-dependent multiplexing. TAM exploits carefully engineered superconducting nanowire geometries to intrinsically encode the spatial location of photon events directly into the detector's electrical output signals. Detection events manifest distinctively in both temporal and amplitude domains of output pulses, enabling simultaneous encoding of spatial information without the need for external circuitry. This innovative encoding strategy effectively overcomes limitations of purely temporal multiplexing methods, particularly timing uncertainty as pixel counts increase. Experimentally validated in a \(32\times32\) pixel kilopixel array, TAM demonstrated high spatial discrimination accuracy (approximately 97\%) and remarkable timing resolution (67.3 ps), while dramatically reducing system complexity and cryogenic wiring demands to just two readout lines. Such dual-domain encoding shows considerable promise for large-area quantum imaging applications, where simplicity and scalability are paramount~\cite{kong2023readout}. Compressive sensing (CS) represents another radical approach that diverges fundamentally from traditional pixel-based wiring schemes. Rather than relying on dedicated wires per pixel, CS employs pseudo-randomized biasing patterns to aggregate photon detection signals across fewer channels. By computationally reconstructing sparse photon event data, CS significantly reduces cryogenic complexity and thermal load. However, successful implementation of CS demands careful consideration of photon flux conditions and computational efficiency in real-time data reconstruction, factors critical for practical applications in quantum optics and remote sensing~\cite{guan2022snspd,luskin2023large}. Thermally-coupled detector architectures provide yet another promising path toward simpler multiplexing. These designs leverage engineered thermal interactions between adjacent superconducting pixels, wherein photon-induced hotspots in one pixel trigger detectable responses in neighboring wires. This thermally mediated signal propagation eliminates extensive electrical interconnects, substantially simplifying wiring without sacrificing detection efficiency. Nevertheless, such systems require meticulous thermal management and optimization to avoid unintended thermal crosstalk or false triggering in densely packed arrays~\cite{allmaras2020demonstration,zhang202332}. Lastly, frequency-division multiplexing (FDM) strategies integrate individual SNSPD pixels with unique superconducting microwave resonators, each tuned to distinct resonance frequencies. Multiple pixels can thus be monitored simultaneously via a single broadband transmission line by detecting subtle frequency shifts or amplitude modulation. This strategy significantly reduces wiring complexity, maintains exceptional timing fidelity, and has been validated successfully in moderate-scale NbN-based arrays. Ongoing research aims to enhance resonator quality factors, minimize inter-channel crosstalk, and extend the applicability of FDM to larger, broadband, multi-spectral detector arrays~\cite{doerner2017frequency,sypkens2024frequency}. Together, these diverse multiplexing and read-out strategies embody significant technological progress toward enabling scalable SNSPD arrays. Each method carries unique strengths and challenges, necessitating continued innovation and cross-disciplinary research in superconducting electronics, cryogenic engineering, computational reconstruction algorithms, and detector architecture optimization. The integration of these advanced multiplexing methods into coherent, practical systems will be crucial for realizing next-generation, large-scale quantum photonic cameras suitable for transformative applications spanning quantum information, biomedical imaging, astronomical sensing, and beyond.

\begin{figure}[htbp]
    \centering
    \includegraphics[width=\linewidth]{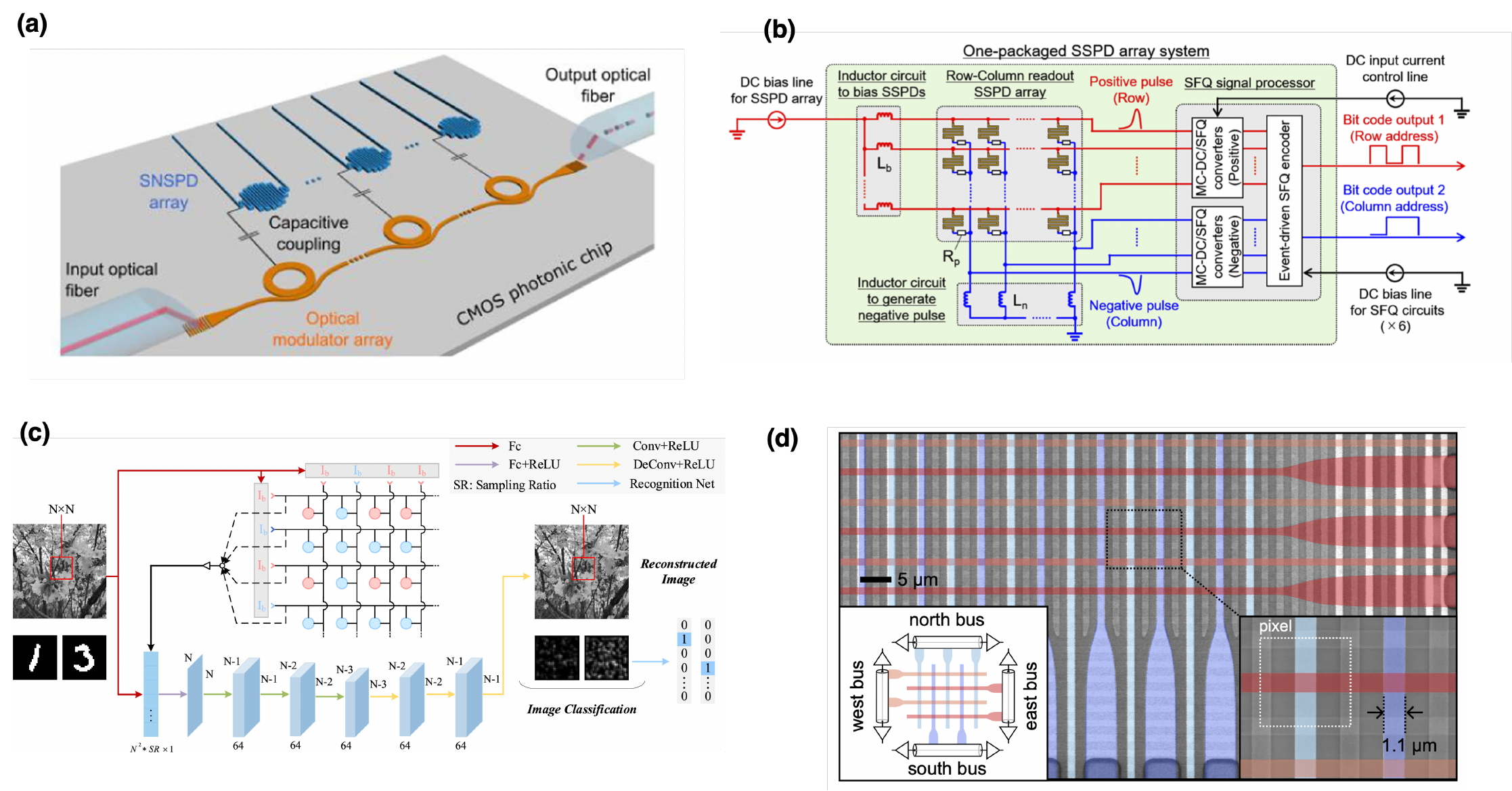}
    \captionsetup{width=\linewidth}
    \caption{Multiplexed Readout Architectures II: Advanced strategies for reading out large-scale superconducting nanowire single-photon detector (SNSPD) arrays.
    (a) Photonic readout of SNSPDs using optical signal encoding to reduce cryogenic complexity and enable scalable high-speed signal extraction~\cite{de2020photonic}.
    (b) Integration of a superconducting digital signal processor for scalable readout of SNSPD arrays using single-flux quantum (SFQ) electronics~\cite{yabuno2020scalable}.
    (c) Compressive sensing-based readout using pseudo-random pixel biasing and computational reconstruction to reduce the number of readout lines~\cite{guan2022snspd}.
    (d) A single-photon camera architecture supporting 400,000 pixels using thermally-coupled imager (TCI) readout and dual-layer SNSPD designs for extreme scalability and polarization insensitivity~\cite{oripov2023superconducting}.
    This TCI approach utilizes asymmetric thermal diffusion to couple detector events into a readout bus, significantly minimizing crosstalk and wiring overhead, while enabling kilopixel to megapixel scalability.
    }
    \label{fig:multiplexing2}
\end{figure}

\section*{Key Challenges in Realizing Scalable SNSPD Imaging Arrays}

 Large-scale SNSPD arrays face increasing challenges related to thermal management, read-out complexity, and maintaining detector uniformity—each of which becomes progressively more severe as array size expands. Thermal management constitutes one of the most critical barriers to SNSPD scalability. SNSPD arrays typically operate under stringent cryogenic conditions, frequently below 4 K and often at sub-Kelvin temperatures, to achieve optimal detection efficiency, minimal dark counts, and picosecond-level timing resolution. At these extreme temperatures, the allowable thermal load is limited, with heat dissipation posing significant risks to system stability and performance\cite{cooper2024optimal,zhang2021investigation}. Individually operated pixels in an SNSPD array require a stable bias current, traditionally supplied from room-temperature electronics via coaxial cables. As the pixel number grows, the cumulative heat conducted through these cables into the cryogenic environment increases dramatically, rapidly surpassing the limited cooling power of standard cryogenic systems. Furthermore, internal sources of dissipation, such as Joule heating from bias resistors and low-noise amplification circuits within the cryogenic stages, magnify this thermal load. The stringent cooling capacity available at temperatures approaching 1 K or below necessitates careful management and mitigation of all parasitic heat sources, complicating system design and potentially limiting practical array sizes. The readout complexity represents another formidable obstacle. Traditionally, each pixel in an SNSPD array requires an independent bias and read-out line, resulting in a quadratic increase in cable count and system complexity as pixel numbers grow. This dense network of coaxial cables not only introduces significant thermal conduction but also leads to substantial mechanical rigidity, complicating cable routing, thermal anchoring, and cryostat design. Furthermore, increased cabling density inevitably raises challenges related to electromagnetic interference (EMI), impedance mismatches, and signal crosstalk—each negatively impacting timing accuracy, signal integrity, and detector fidelity. Maintaining sub-100-ps timing resolution across hundreds or thousands of pixels necessitates precise control of the connections, a challenge that becomes increasingly difficult with extensive wiring infrastructure and longer cable lengths. Recent advancements in photonic readout of SNSPDs provide an attractive alternative to overcome some of these issues\cite{de2020photonic,thiele2023all}. On a more fundamental level, the physics governing SNSPD operation imposes additional constraints when scaling detector arrays. High-density arrays must maintain precise uniformity in superconducting film thickness, composition, and performance \cite{zichi2019optimizing} to ensure consistent bias current distribution and uniform photon detection efficiency across all pixels. Variations in fabrication parameters, material quality, or external contamination can lead to significant discrepancies in critical current thresholds, detection efficiency, and timing jitter between pixels, adversely affecting overall array performance and reliability\cite{wollman2019kilopixel}. Moreover, at high pixel densities, thermal and electromagnetic interactions between adjacent pixels become critical. Localized heating caused by a photon detection event in one pixel may inadvertently propagate thermally or electrically to neighboring pixels, inducing unwanted detector counts or lower temporal resolution. Such pixel-to-pixel interactions severely limit the achievable spatial and temporal resolution of closely packed detector arrays, necessitating innovative approaches to manage or mitigate these cross-talk mechanisms. Ultimately, overcoming these intertwined physical and engineering challenges demands a holistic rethinking of SNSPD array design and integration strategies.  Additionally, signal processing becomes a significant bottleneck as the number of pixels in SNSPD arrays increases. The burden of accurately time-tagging a high volume of photon detection events grows rapidly, especially in systems operating at gigahertz count rates across thousands of detectors. Each detection event must be timestamped with picosecond precision to preserve the temporal resolution essential for applications such as quantum communication, time-of-flight imaging, and time-correlated single-photon counting (TCSPC). This requires fast, low-jitter electronics with high bandwidth and minimal latency. Moreover, implementing high-order temporal corrections, such as compensating for detector-specific timing offsets (e.g., due to variations in kinetic inductance or routing delays), becomes increasingly complex in large arrays. Calibration routines must account for pixel-to-pixel timing mismatches and synchronization errors across distributed cryogenic and room-temperature electronics. As system sizes grow, so does the need for scalable, parallelizable digital signal processing pipelines capable of handling massive data throughput in real time. This may involve field-programmable gate arrays (FPGAs), custom application-specific integrated circuits (ASICs), or cryo-compatible digital logic integrated closer to the detector plane.  Addressing these issues will require innovations spanning multiple domains, including superconducting and substrate materials engineering, cryogenic electronics, advanced multiplexed read-out architectures, and robust cryogenic packaging solutions. Only through such interdisciplinary efforts can the full potential of scalable SNSPD arrays be realized, enabling robust single-photon cameras capable of transforming photon-limited imaging applications across quantum technologies, biomedical diagnostics, remote sensing, and astronomy.

\section*{Roadmap and Vision for Next-Generation SNSPD Arrays}

To realize the transformative potential of superconducting nanowire single-photon detectors (SNSPDs) as practical quantum imaging cameras, a strategic roadmap that integrates innovations across materials science, device engineering, multiplexing strategies, and system-level cryogenic integration is essential. Although recent developments have showcased significant breakthroughs in kilopixel-scale arrays and novel readout approaches, substantial challenges and opportunities remain in achieving fully integrated, robust, and large-format SNSPD imaging systems. At the device level, immediate progress requires improving the uniformity and reproducibility of detector performance across centimeter- to wafer-scale arrays. Achieving arrays with thousands of pixels necessitates significant improvements in superconducting film homogeneity, precise control over critical parameters such as nanowire width, thickness, and kinetic inductance, and optimization of lithographic fabrication processes. Establishing foundry-compatible processes employing scalable techniques such as deep ultraviolet lithography and standardized film deposition methods will ensure reproducible fabrication with minimal variability, laying the groundwork for reliable, high-yield manufacturing \cite{psiquantum2025manufacturable}. Addressing the critical bottleneck of cryogenic system integration is equally vital. While existing multiplexing solutions such as row-column readout, frequency-domain multiplexing, and superconducting single-flux quantum (SFQ) electronics have demonstrated feasibility at moderate scales, their full integration into commercial-scale systems remains to be accomplished. Developing hybrid cryogenic electronics that integrate digital processing directly within the detector chip is a promising route toward substantial reductions in thermal load and cabling complexity. Leveraging emerging cryogenic CMOS and mixed-signal technologies designed explicitly for operation at sub-Kelvin temperatures can offer unprecedented scalability by minimizing power dissipation, cable count, and electromagnetic interference. In parallel, novel multiplexing concepts such as orthogonal time–amplitude multiplexing, compressive sensing-based reconstruction, and advanced hybrid AQFP/RSFQ logic interfaces represent exciting directions for radical improvement in array scalability. These strategies shift the complexity from hardware-dominated wiring and analog electronics to algorithmically-driven reconstruction and minimalistic hardware architectures. By adopting these computationally assisted readout methods, future SNSPD arrays could scale to tens or even hundreds of thousands of pixels with remarkably reduced cryogenic overhead, enabling compact, power-efficient, and robust photon-counting cameras. Thermal management remains one of the most fundamental challenges that must be continuously addressed through innovation in materials engineering and cryostat design. The implementation of advanced passive cooling materials, improved thermal anchoring techniques, and integrated cryogenic cooling solutions tailored specifically to SNSPD arrays will be critical. Additionally, exploring novel superconducting materials with higher critical temperatures or lower heat dissipation could significantly relax stringent thermal constraints, allowing operation at higher cryogenic temperatures with greater reliability and practicality. At the system and integration levels, significant opportunities lie in optical engineering to efficiently couple photons into SNSPD arrays. Advances in microlens arrays, metamaterials, integrated photonic cavities, and 3D-printed optical structures must be leveraged to maximize photon collection efficiency and uniformity across large detector areas. Incorporating reconfigurable and dynamically tunable optical elements can further enable adaptive imaging systems optimized for specific quantum sensing or biomedical imaging scenarios. Looking forward, the ultimate vision involves fully integrated, monolithic SNSPD imaging chips comprising tens of thousands of pixels, seamlessly interfaced with compact, low-power cryogenic electronics and adaptive optical elements. To support such large-scale systems, a dedicated software infrastructure for imaging reconstruction, event correlation, and real-time analysis will also be essential. Such systems could incorporate signal processing using neuromorphic algorithms or machine learning-based reconstruction techniques to facilitate real-time, high-fidelity imaging in photon-starved conditions. Achieving this ambitious goal will require sustained interdisciplinary collaboration involving quantum physicists, electrical engineers, materials scientists, optical designers, and computational imaging experts. By systematically addressing these areas of materials uniformity, cryogenic integration, multiplexing innovation, thermal management, optical coupling, and computational reconstruction, we anticipate transformative impacts across quantum communications, photonic quantum computing, deep-tissue biomedical imaging, and next-generation space-based sensing. The proposed roadmap is ambitious, yet achievable, provided focused investments in fundamental research, interdisciplinary collaboration, and technology standardization are prioritized, ultimately positioning SNSPD technology at the forefront of the quantum and photonics revolution.

\section*{Emerging Opportunities Enabled by SNSPD Array Technology}
The realization of scalable superconducting nanowire single-photon detectors (SNSPD) arrays represents a landmark development poised to revolutionize numerous scientific and technological fields by overcoming fundamental limits of photon detection. With exceptional detection efficiency, ultralow noise, precise timing capabilities, and robust multiplexed architectures, large-format SNSPD cameras promise unprecedented capabilities in quantum science, biomedical imaging, advanced sensing, and classical communication applications. In the domain of biomedical imaging, scalable SNSPD arrays uniquely address critical bottlenecks in photon-starved modalities such as fluorescence lifetime imaging microscopy (FLIM) and Raman spectroscopy. These techniques critically depend on precise photon arrival timing and extremely low background counts to resolve intricate biological processes at the single-molecule and cellular levels\cite{chen2017mid,buschmann2023integration,wang2022vivo}. SNSPDs, with their picosecond-scale jitter and near-perfect quantum efficiency, substantially enhance the spatial and temporal resolution, significantly improving diagnostic capabilities in cancer imaging, intracellular dynamics studies, and neuroscience. Furthermore, the inherent broadband detection capabilities of SNSPDs extending into the near-infrared spectral range dramatically improve the penetration depth for deep-tissue imaging methods like diffuse optical tomography (DOT) and photoacoustic tomography, providing vital tools for non-invasive medical diagnostics and therapy\cite{xia2021short,parfentyeva2023fast} monitoring. Quantum information science stands to benefit profoundly from scalable SNSPD arrays\cite{beutel2021detector}. High-performance arrays form the cornerstone of next-generation quantum networks, enabling robust and efficient quantum key distribution (QKD), entanglement swapping, and quantum repeater technologies necessary for global quantum communication infrastructure. Additionally, SNSPD arrays are indispensable in photonic quantum computing architectures, particularly linear optical quantum computing and boson sampling schemes, which rely on precise photon-number-resolving capabilities, ultra-low photon loss, and rapid detection rates. By providing fast, reliable photon counting with minimal latency and high throughput, scalable SNSPD arrays enable critical advancements toward fault-tolerant quantum processors and scalable quantum internet platforms\cite{psiquantum2025manufacturable}. 

The emergence of quantum sensing and imaging modalities such as quantum ghost imaging, quantum illumination, and quantum-enhanced metrology also hinges upon large-scale SNSPD technology. These quantum-enabled techniques exploit photon correlations and entanglement, allowing imaging and sensing beyond classical shot-noise limitations, potentially revolutionizing remote sensing, environmental monitoring, and security imaging. Classical technologies such as autonomous navigation and remote sensing further benefit significantly from SNSPD-based LiDAR systems\cite{guan2022lidar}. With unmatched timing accuracy, exceptional photon sensitivity, and the ability to discriminate weak signals from intense daylight backgrounds, SNSPD arrays offer robust and high-resolution environmental sensing critical for self-driving vehicles, airborne mapping, and advanced drone systems. Space exploration and astronomy represent further critical application domains. SNSPD arrays enable direct photon-counting detectors capable of capturing extremely faint signals from distant astronomical sources with minimal background noise and high timing precision. Their deployment in future space telescopes and planetary observation platforms would enable groundbreaking discoveries in exoplanet imaging, transient event detection, and dark-matter detection \cite{bass2024quantum,hochberg2022new}.  Finally, scalable SNSPD arrays are paving the way for emerging interdisciplinary applications, including neuromorphic photonic computing\cite{buckley2020integrated} and quantum-enhanced optical coherence tomography (OCT)\cite{defienne2024advances}. The convergence of SNSPD technology with integrated photonics, cryogenic electronics, and advanced computational imaging methodologies promises transformative breakthroughs, far surpassing the capabilities achievable with conventional semiconductor photodetectors. Ultimately, scalable SNSPD arrays are not incremental improvements but rather represent a foundational shift in photon detection technology, enabling revolutionary capabilities across quantum and classical photonics, biomedical science, astronomy, remote sensing, and beyond. 

\begin{figure}[htbp]
    \centering
    \includegraphics[width=\linewidth]{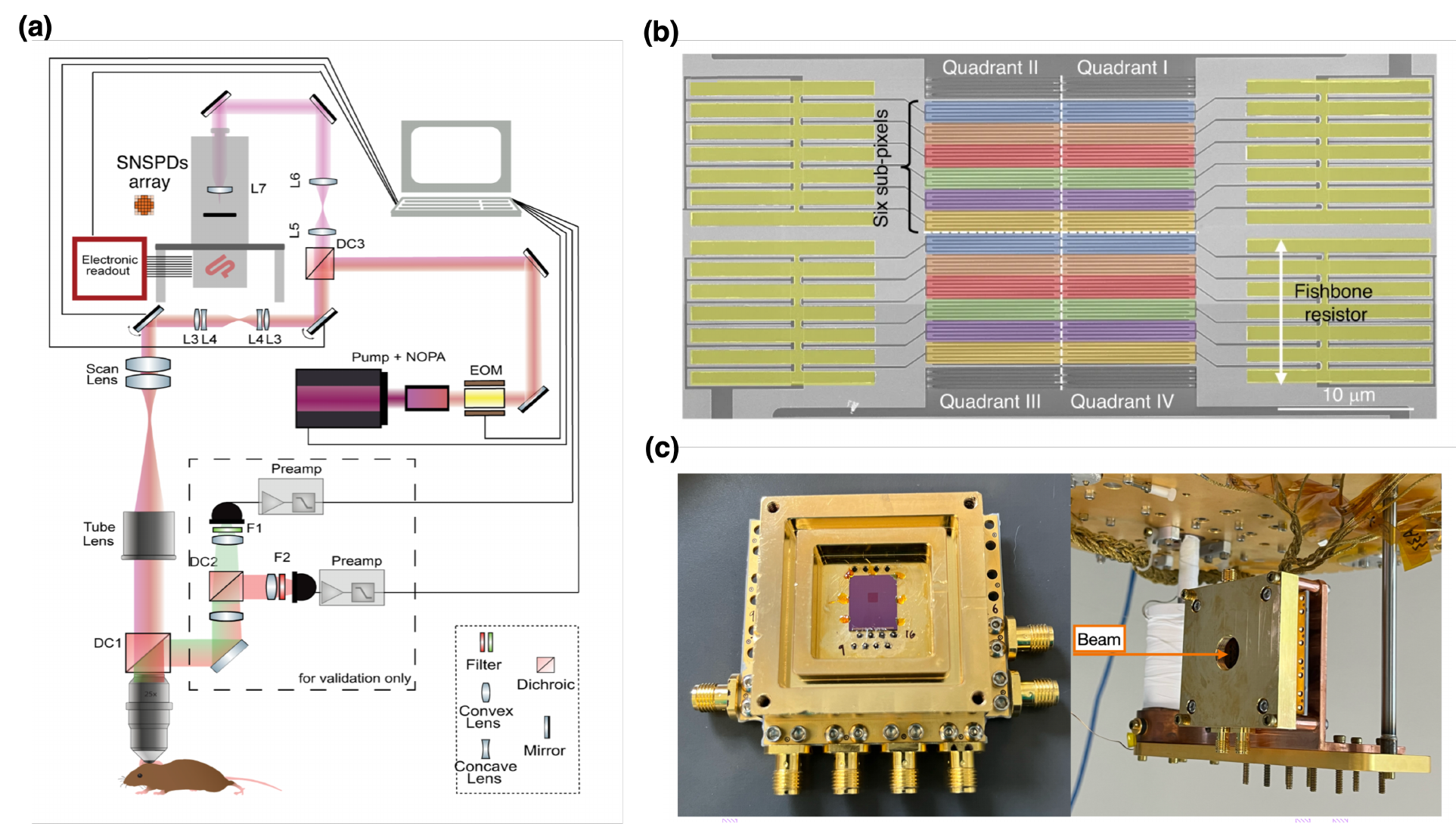} 
    \caption{Representative applications enabled by scalable SNSPD arrays.
    (a) Deep-tissue near-infrared two-photon fluorescence imaging of the mouse brain using a superconducting nanowire single-photon detector (SNSPD) array demonstrates superior sensitivity and depth penetration \cite{tamimi2024deep}.
    (b) A compact multi-pixel SNSPD system enabling gigabit space-to-ground optical communications, illustrating the potential of SNSPD arrays in high-speed quantum-secure communication links \cite{hao2024compact}.
    (c) Large-area superconducting microwire arrays used for high-energy particle detection, highlighting the scalability and robustness of SNSPDs for use in fundamental physics and astroparticle experiments \cite{pena2025high}.
    }
    \label{fig:snsdp-array-architecture}
\end{figure}

\section*{Outlook and Future Directions}

We are at a pivotal moment in the development of scalable superconducting nanowire single-photon detectors (SNSPDs). Decades of fundamental research and incremental innovations have brought SNSPDs from laboratory curiosities to the forefront of candidates for next-generation photon detection technologies. The path forward, however, demands a coordinated interdisciplinary effort encompassing materials science, cryogenic electronics, computational imaging, quantum optics, and systems engineering. Addressing key scalability bottlenecks—such as thermal load management, cryogenic wiring complexity, device uniformity, and system-level integration—is paramount. Continued development of advanced multiplexing techniques such as time–amplitude multiplexing, hybrid superconducting logic architectures, and compressive sensing read-outs will play critical roles in overcoming these challenges. Simultaneously, breakthroughs in superconducting material engineering, lithographic precision, and scalable nanofabrication processes will enable consistent and uniform detector arrays necessary for practical deployment.  The integration of SNSPD arrays with innovative optical coupling techniques—leveraging advances in microlenses, metasurfaces, and photonic crystals—will further enhance system-level efficiency, reliability, and adaptability. Moreover, the adoption of advanced computational methods, such as machine learning and neuromorphic algorithms for photon-event reconstruction and real-time image processing, will unlock new dimensions in array scalability and performance, bridging hardware constraints with computational flexibility. Realizing the full potential of SNSPD arrays as a transformative technology will require close collaboration between academia, industry, and government institutions. The standardization of fabrication techniques, the development of scalable manufacturing processes, and the establishment of integrated cryogenic platforms will accelerate the transition from laboratory prototypes to commercial products and practical systems. Beyond improved sensitivity and scalability, SNSPD cameras offer new opportunities to extend the limits of quantum measurement. For example, they may enable the real-time reconstruction of single-photon quantum events with few-picosecond and multi-pixel resolution, providing access to spatiotemporal dynamics beyond current single-pixel detection capabilities. At larger scales, such arrays could support quantum-resolved cosmological imaging, detecting individual relic photons with spectral and polarization detail, and opening new avenues for probing early-universe physics. These emerging directions illustrate the broader potential of SNSPD arrays as platforms for fundamentally new types of sensing and discovery. In conclusion, scalable SNSPD arrays promise a significant leap in quantum detection capabilities, opening entirely new avenues for fundamental science and innovative technologies. Their continued development is not just an incremental step but a strategic investment in a future defined by quantum-enhanced photonics, ultra-sensitive imaging, and transformative advancements across multiple scientific and technological disciplines. The next decade will be crucial in establishing SNSPD arrays as ubiquitous tools at the forefront of the quantum photonic revolution, profoundly impacting science, technology, quantum systems industry, and society at large.

\section*{Declarations}

\subsection*{Availability of data and materials}
Not applicable.

\subsection*{Competing interests}
The authors declare that they have no competing interests.

\subsection*{Funding}
J.G. acknowledges support from Swedish Research Council (Ref: 2023-06671 and 2023-05288), Vinnova project (Ref: 2024-00466) and the Göran Gustafsson Foundation. A.W.E acknowledges support from Knut and Alice Wallenberg (KAW) Foundation through the Wallenberg Centre for Quantum Technology (WACQT), Swedish Research Council (VR) Starting Grant (Ref: 2016-03905). V.Z. acknowledges support from the KAW.

\subsection*{Authors' contributions}
All authors contributed to the conception, design, and writing of the manuscript. All authors read and approved the final version.


\end{document}